\documentclass[11pt,a4paper]{article}
\usepackage{jheppub}
\usepackage[T1]{fontenc}

\allowdisplaybreaks

\newcommand{\0}{SO(10)}

\newcommand{\Group}{$\textrm{SU(3)}_C\times\textrm{SU(2)}_L\times \textrm{U(1)}_Y$}

\newcommand{\be}{\begin{equation}}
\newcommand{\ee}{\end{equation}}
\newcommand{\bea}{\begin{eqnarray}}
\newcommand{\eea}{\end{eqnarray}}

\newcommand{\pht}[1]{\Phi_{#1}}

\title{Connection between proton decay suppression and seesaw mechanism in supersymmetric SO(10) models}

\author{Lipei Du,}
\author{Xiaojia Li}
\author{and Da-Xin Zhang}
\affiliation{School of Physics and State Key Laboratory of Nuclear Physics and Technology, \\
Peking University, Beijing 100871, China}
\emailAdd{lpdu@pku.edu.cn}
\emailAdd{shakalee@pku.edu.cn}
\emailAdd{dxzhang@pku.edu.cn}

\abstract{
We propose a mechanism to suppress proton decay induced by
dimension-5 operators in a supersymmetric SO(10) model.
Proton lifetime is directly connected with the  intermediate  vacuum expectation value
which is responsible for the seesaw mechanism.
The model shows many consistencies with the present theoretical results such as
the components of the two Higgs doublets in the minimal supersymmetric standard model.}

\begin{document}
\maketitle
\flushbottom
\pagenumbering{arabic}
\section{Introduction}\label{inttro}
The Grand Unified Theory (GUT) \cite{gut1,gut2} has never failed to fascinate
the particle physicists since it was proposed.
Besides unifying the gauge interactions in a simple group,
it also unifies  quarks and leptons in same multiplets which, however, are
very different in \Group{} of the standard model (SM).
Consequently, lepton and baryon numbers are not conserved separately.
While lepton number violation is strongly supported by the observation
of neutrino oscillations  which are usually explained through the seesaw mechanism \cite{seesawi1,seesawi2,seesawi3,seesawi4,seesawi5,seesawii1,seesawii2,seesawii3,seesawii4},
the baryon number violation is strongly constrained by the proton decay experiments
which need a natural explanation.

Supersymmetric (SUSY) GUT model based on SO(10) \cite{so101,so102}, among the various GUT models,
is very attractive due to its several advantages.
Firstly, as protected by supersymmetry,
it has no problem in naturalness.
Secondly, having all the fermions of a generation contained in one \textbf{16} dimensional representation which contains
the right-handed neutrino,
the model can naturally explain the neutrino oscillations through the
seesaw mechanism.
Thirdly, in the minimal \cite{msso10a1,msso10a2,msso10b} and the next-to-minimal \cite{nmsso101,nmsso102} versions of SUSY SO(10),
the theories are renormalizable and $R$-parity is conserved which
prohibits the most dangerous dimension-4 operators for proton decay.

Nevertheless, the SUSY SO(10) models have also difficulties to overcome.
To realize the seesaw mechanism, an intermediate scale is usually introduced.
This will generally bring in new particles at this scale and break down gauge coupling
unification badly \cite{msso10b}.
However, as noticed recently in \cite{dlz}, there is actually no need to introduce an intermediate seesaw scale above which new gauge interactions begin.
Instead, the seesaw mechanism requires only an intermediate vacuum expectation value (VEV) which contributes only a small portion of the new gauge bosons whose masses are still around the GUT scale.
As a consequence, the gauge coupling unification will not be broken down as in those models, e.g., the minimal SUSY SO(10) model (MSSO10).

Furthermore, in the SUSY models, the dimension-5 operators dominate the proton decay rates and therefore strongly need to be suppressed by a mechanism.
In the literature, since these operators are related to the Yukawa
couplings, careful adjustments of the Yukawa couplings \cite{moha} are common which however are not sufficient
as the lower limit on the proton lifetime from experiments is increasing.

In this work, instead of strictly solving the doublet-triplet splitting problem  labored by many groups \cite{dts1,dts2},
we simply assume that the Higgs doublet pair in the minimal supersymmetric standard model (MSSM)
are achieved by fine-tuning which will not be performed explicitly.
Our efforts are mainly focused on proposing a mechanism to sufficiently suppress the proton decay rates.
We will  extend the MSSO10 to achieve this goal.
The effective triplet mass (ETM) \cite{ETM1,ETM2},
to which the dimension-5 operators are inversely proportional,
is enhanced due to the special structure of the color-triplet Higgs mass matrix.
This suppression of proton decay is found to be directly related to the
intermediate VEV required by the seesaw mechanism.
We also find that the massless  MSSM doublets obtained by the assumed fine-tuning
are also related to the intermediate VEV,
and that these doublets conform to the results from simply fitting the fermion sector in SO(10) models without considering other stringent constraints.

In the next section we will present the model,
followed by the realization of the seesaw mechanism in Section \ref{seesawsec}.
The solution of the model required by SUSY is presented in Section \ref{susypre}.
The mechanism of suppression proton decay follows in Section \ref{tripletproton}.
Predictions on the MSSM Higgs doublets are given in Section \ref{doubletsec}.
We will summarize finally.


\section{The present model}\label{present}
The present model contains the following particles in the spectrum.
First, each generation of the matter superfields  are contained in a \textbf{16}-plet superfields $\psi_i${} $(i=1, 2, 3)${} as in most of the SO(10) models.
Second, we use \textbf{210}-plet Higgs to break GUT symmetry.
To further break $\textrm{U(1)}_R\times\textrm{U(1)}_{B-L}$ symmetry down to U(1)$_Y$, two pairs of \textbf{126}+$\overline{\textbf{126}}$-plet Higgs (denoted by $\Delta_i+\overline{\Delta}_i, i=1,2$) are introduced.
Two Higgs doublets in \textbf{10} ($H_{1,2}$), together with those in the \textbf{126}+$\overline{\textbf{126}}$s, are used to break down the electroweak symmetry.
Third, we will introduce a U(1) symmetry to differentiate these Higgs into
those couple with the matter fields and those  do not.
These  U(1) quantum numbers $Q$ are listed in Table \ref{Qnumbers}.
\begin{table}[!htp]
\begin{center}
\caption{\label{Qnumbers}\0 multiplets and their U(1) charges.}
\begin{tabular}{cccccccc}
\hline
\hline
Charges & $\psi_{i}$ & $H_1$ & $\Delta_1/\overline{\Delta}_1$ & $H_2$ & $\Delta_2/\overline{\Delta}_2 $ & $\Phi$ & $S$\\
\hline
$Q$ &$-1/2$ &1 &1 &-1  &-1 &0 &2\\
\hline
\hline
\end{tabular}
\end{center}
\end{table}

Here we will simply treat the U(1) symmetry as a global one broken by the VEV of a SO(10) singlet $S$
which is taken as
\begin{equation}\label{S0}
S_0\sim M_I\sim 10^{14}\textrm{GeV}\sim 10^{-2}M_{G}.
\end{equation}
In Section \ref{seesawsec} this VEV $S_0$ will naturally generate the seesaw VEV and thus the model has no mass larger than the GUT scale explicitly.
The value $S_0$ in (\ref{S0}) is also of the order of $\frac{M_G^2}{M_{Plank}}$,
 which may suggest alternatively that it is possible to be realized through an analogue of a seesaw mechanism,
if we treat the U(1) as an anomalous symmetry broken by a Planck scale VEV generated
by the Green-Schwarz mechanism \cite{green1,green2,green3,green4}.
For simplicity, this later possibility will not be discussed further.

The matter fields are negative in U(1) charges, so the Yukawa superpotential is
\begin{equation}\label{superp1}
W_Y=Y_{10}^{ij}\psi_i\psi_j H_1+Y_{126}^{ij}\psi_i\psi_j \overline{\Delta}_1,
\end{equation}
which is just the same as in the MSSO10. The most general renormalizable superpotential in the Higgs sector is given by
\begin{eqnarray}  \label{superp2}
W_H&=&\frac{1}{2}m_{\Phi}\Phi^2+{m_{\Delta}}_{12}\overline{\Delta}_1\Delta_2
+{m_{\Delta}}_{21}\overline{\Delta}_2\Delta_1+m_{H}H_1H_2 \nonumber \\
&+&(\beta_{12}\Delta_1+\overline{\beta}_{12}\overline{\Delta_1})H_2\Phi+(\beta_{21}\Delta_2
+\overline{\beta}_{21}\overline{\Delta_2})H_1\Phi\nonumber \\
&+&\lambda\Phi^3+(\lambda_{12}\overline{\Delta}_1\Delta_2 +{\lambda}_{21}\overline{\Delta}_2\Delta_1)\Phi+S(\frac{1}{2}\alpha_1H_2^2+\alpha_2\overline{\Delta}_2\Delta_2).
\end{eqnarray}


\section{On the seesaw mechanism}\label{seesawsec}
The small but non-vanishing neutrino masses can be naturally explained using the seesaw mechanism.
In a model where the type-I seesaw dominates,
the mass matrix of neutrinos is given as $M_{\nu}\simeq-M_{\nu_D}^T M_{\nu_R}^{-1} M_{\nu_D}$.
The Majorana mass matrix $ M_{\nu_R}$ comes from the VEV of a SU(2)$_R$ triplet
 contained in $\overline{\textbf{126}}$, which corresponds to the seesaw scale $M_I$.
A sub-eV neutrino mass roughly indicates $M_I\sim 10^{14}\textrm{GeV}\sim 10^{-2}M_{G}$.
However, the presence of an intermediate scale breaks the unification of gauge couplings badly \cite{msso10b} in general.

In the present model, the presence of two $\overline{\textbf{126}}$s changes the situation and the GUT symmetry would be broken down to the SM symmetry directly.
Instead of an intermediate scale, only an intermediate valued  VEV, i.e. $\overline{v}_{1R}$, of the order $O(M_I)$ is required to
couple with the matter fields \cite{dlz}.
The D-flatness required by SUSY at high energy scales is
\begin{equation}  \label{dterm}
|v_{1R}|^2+|v_{2R}|^2=|\overline{v}_{1R}|^2+|\overline{v}_{2R}|^2,
\end{equation}
where the $v$s and $\overline{v}$s are the VEVs of
the SU(2)$_R$ triplets in \textbf{126}s and $\overline{\textbf{126}}$s, respectively.
Eq. (\ref{dterm}) can be fulfilled even if $\overline{v}_{1R}$ is small compared to the GUT scale.
Then the seesaw mechanism does not conflict with gauge coupling unification if the other VEVs are taken at the GUT scale.


\section{SUSY preserving at high energy}\label{susypre}
When the \0 breaks down to the MSSM,  only the MSSM singlets can get VEVs,
\begin{eqnarray}\label{allvev}
\Phi _{1}&=&\langle \Phi (1,1,1)\rangle, {}~\Phi _{2}=\langle \Phi
(15,1,1)\rangle,{}~\Phi _{3}=\langle \Phi (15,1,3)\rangle;  \nonumber \\
v_{(1,2)R}&=&\langle \Delta_{(1,2)} (\overline{10},1,3)\rangle,{}~\overline{v}_{(1,2)R}
=\langle\overline{\Delta}_{(1,2)}(10,1,3)\rangle.
\end{eqnarray}
The Pati-Salam ($SU(4)_C\times SU(2)_L \times SU(2)_R$) subgroup indices are used to specify different singlets of the MSSM. Substituting  these VEVs into (\ref{superp2}), we get
\begin{eqnarray}  \label{superp2VEV}
\langle W_H \rangle &=&\frac{1}{2}m_{\Phi}(\Phi_1^2+\Phi_2^2+\Phi_3^2)+\lambda(\frac{1}{9\sqrt{2}}\Phi _{2}^{3}+\frac{1}{2\sqrt{6}}\Phi _{1}\Phi _{3}^{2}+\frac{1}{3\sqrt{2}}\Phi _{2}\Phi _{3}^{2})+{m_{\Delta}}_{12}\overline{v}_{1R}v_{2R}\nonumber\\
&+&{m_{\Delta}}_{21}\overline{v}_{2R}v_{1R}
+(\lambda_{12}\overline{v}_{1R}v_{2R} +{\lambda}_{21}\overline{v}_{2R}v_{1R})\Phi_0+\alpha_2 S_0 \overline{v}_{2R}v_{2R},
\end{eqnarray}
where we have defined
$$\Phi_0=\left[\Phi_1\frac{1}{10\sqrt6}+\Phi_2\frac{1}{10\sqrt2}+\Phi_3\frac{1}{10}\right].$$

In the presence of all the VEVs in (\ref{allvev}), to preserve SUSY at high energy, besides the D-flatness condition in (\ref{dterm}),
the F-flatness conditions are also required, i.e.,
\begin{equation}
\left\{ \frac{\partial }{\partial \Phi _{1}},\frac{\partial }{\partial \Phi
_{2}},\frac{\partial }{\partial \Phi _{3}},\frac{\partial }{\partial v_{1R}},
\frac{\partial }{\partial \overline{v}_{1R}},\frac{\partial }{\partial v_{2R}},
\frac{\partial }{\partial \overline{v}_{2R}}\right\} \langle W_H\rangle =0.
\label{partiale}
\end{equation}
Then we get
\begin{eqnarray}
0&=&m_{\Phi}\Phi_1+\frac{\lambda\Phi_3^2}{2\sqrt6}+\frac{1}{10\sqrt6}(\lambda_{12}\overline{v}_{1R}v_{2R} +{\lambda}_{21}\overline{v}_{2R}v_{1R}),  \nonumber\\
0&=&m_{\Phi}\Phi_2+\frac{\lambda\Phi_2^2}{3\sqrt2}+\frac{\lambda\Phi_3^2}{3\sqrt2}+\frac{1}{10\sqrt{2}}(\lambda_{12}\overline{v}_{1R}v_{2R} +{\lambda}_{21}\overline{v}_{2R}v_{1R}), \label{equphi} \\
0&=&m_{\Phi}\Phi_3+\frac{\lambda\Phi_1\Phi_3}{\sqrt6}+\frac{\sqrt2\lambda\Phi_2\Phi_3}{3}+\frac{1}{10}(\lambda_{12}\overline{v}_{1R}v_{2R} +{\lambda}_{21}\overline{v}_{2R}v_{1R}), \nonumber
\end{eqnarray}
for $\Phi _{1,2,3}$, respectively. The condition for $v_{1R}$ and ${v}_{2R}$ is
\begin{equation}
\left(
\begin{array}{cc}
\overline{v}_{1R} & \overline{v}_{2R}
\end{array}
\right)
\left(
\begin{array}{cc}
0 &M_{12} \\
M_{21}& \alpha_2 S_0
\end{array}
\right)
=0, \label{equvr1} \\
\end{equation}
and that for $\overline{v}_{1R}$ and $\overline{v}_{2R}$ is
\begin{equation}
\left(
\begin{array}{cc}
0 &M_{12} \\
M_{21}& \alpha_2 S_0
\end{array}
\right)
\left(
\begin{array}{c}
{v}_{1R}\\
{v}_{2R}
\end{array}
\right)=0. \label{equvr2} \\
\end{equation}
Here for simplicity we defined
\begin{equation}
M_{12}={{m}_{\Delta}}_{12}+{\lambda}_{12}\Phi_0 \quad M_{21}={{m}_{\Delta}}_{21}+{\lambda}_{21}\Phi_0. \label{M1221} \\
\end{equation}

Equations (\ref{equvr1}) and  (\ref{equvr2}) both require
\begin{equation}
\textrm{Det}\left(
\begin{array}{cc}
0 &M_{12} \\
M_{21}& \alpha_2 S_0
\end{array}
\right)=M_{12}M_{21}=0. \label{detv} \\
\end{equation}
If we check the mass matrix of the SM singlets,
we can see that under (\ref{detv}) the massless Goldstone mode responsible for $U(1)_{I_{3R}}\times U(1)_{B-L} \to U(1)_Y$ can be generated while all the other eigenstates in the same SM representation remain massive.

If we take $M_{21}=0$, we can get the following solutions
\begin{equation}
-\frac{\overline{v}_{1R}}{\overline{v}_{2R}}=\frac{\alpha_2 S_0}{M_{12}}\sim 10^{-2},\quad {v}_{2R}=0, \label{M1221} \\
\end{equation}
which means we can naturally get the seesaw VEV $\overline{v}_{1R}$ by considering the F-flatness conditions because of the intermediate VEV $S_0$. It is not that both $\overline{v}_{1R}$ and $\overline{v}_{2R}$ get independent VEV, but only a combination of them gets VEV whose main component comes from $\overline{v}_{2R}$.
For a vanishing $v_{2R}$,
we define $\Phi_3=6 m_{\Phi}x/\lambda$ following \cite{msso10b} and get
\begin{eqnarray}
\Phi_1&=&-\frac{\sqrt6 m_{\Phi}}{\lambda}\frac{x(1-5x^2)}{(1-x)^2}, \nonumber \\
\Phi_2&=&-\frac{3\sqrt2 m_{\Phi}}{\lambda}\frac{(1-2x-x^2)}{(1-x)}, \label{solvrvrbar}\\
\lambda_{21}\overline{v}_{2R}v_{1R}&=&\frac{60m_{\Phi}^2}{\lambda}\frac{x(1-3x)(1+x^2)}{(1-x)^2}.\nonumber
\end{eqnarray}
The $x$ is then determined by $M_{21}=0$ and thus determines $\overline{v}_{2R}\sim v_{1R}$ which are generally
at the GUT scale.

If a vanishing $M_{12}$ after (\ref{detv}) is taken instead, we can not get the wanted seesaw VEV and the further results of fermion masses are inconsistent with experiments. For these reasons, the $M_{12}=0$ case will not be discussed further below.

In summary, SUSY at high energy and the seesaw mechanism choose to satisfy
\begin{equation}
\overline{v}_{1R}=M_I, ~v_{2R}=0,
\end{equation}
for the SO(10) symmetry breaking and thus
\begin{equation}\label{v1rv2rbar}
|\overline{v}_{2R}|\sim |v_{1R}|=\sqrt{\left|\frac{60m_{\Phi}^2}{\lambda\lambda_{21}}\frac{x(1-3x)(1+x^2)}{(1-x)^2}\right|}
\end{equation}
following (\ref{solvrvrbar}).
All Higgs superfields are given masses  at the GUT scale except the two
doublets in MSSM whose masses  require a minimal fine-tuning of the parameters as done
in the MSSO10 \cite{msso10b}.
Then gauge coupling unification can be realized by adjusting other parameters
of the model.


\section{The triplet mass matrix and suppression of proton decay}\label{tripletproton}
All the Higgs multiplets in Table \ref{Qnumbers} contain color triplet-antitriplet pairs.
The color triplets are ordered as
\begin{equation}\label{phit}
\varphi_{T}=(H_{1T},\Delta_{1T}, \overline{\Delta}_{1T}, \overline{\Delta}'_{1T}, \Phi_{T}, H_{2T},\Delta_{2T}, \overline{\Delta}_{2T}, \overline{\Delta}'_{2T}),
\end{equation}
while the color antitriplets are
\begin{equation}\label{antiphit}
\varphi_{\overline{T}}=(H_{1\overline{T}}, \overline{\Delta}_{1\overline{T}}, \Delta_{1\overline{T}}, \Delta'_{1\overline{T}}, \Phi_{\overline{T}}, H_{2\overline{T}}, \overline{\Delta}_{2\overline{T}}, \Delta_{2\overline{T}}, \Delta'_{2\overline{T}}).
\end{equation}
The mass term of the Higgs color triplets is  given by $(\varphi_{\overline{T}})_a(M_T)_{ab}(\varphi_T)_b$,
 with the $9\times 9$ matrix $M_T$ written as
\begin{equation}\label{triplet}
M_T=\left(
\begin{array}{cc}
B_{11(4\times 4)} &B_{12(4\times 5)} \\
B_{21(5\times 4)}& B_{22(5\times 5)}
\end{array}
\right).
\end{equation}
The $B_{11}$ is a $4\times 4$ null matrix, and the rests are \cite{nmsso101,nmsso102}
\begin{equation}\label{MT12}
B_{12}=\left(
\begin{array}{ccccc}
\frac{\overline{\beta}_{21}\overline{v}_{2R}}{\sqrt{5}} & m_{H} & \beta_{21}\pht{H\Delta} & \overline{\beta}_{21}\pht{H\overline{\Delta}} &\frac{-\sqrt{2}~\overline{\beta}_{21}\Phi_{3}}{\sqrt{15}}   \\
0 & \overline{\beta}_{12}\pht{H{\Delta}} & {m}_{\Delta_{12}} & 0 & 0\\
    \frac{-{\lambda}_{21}\overline{v}_{2R}}{10\sqrt{3}} & \beta_{12}\pht{H\overline{\Delta}} & 0 & {m}_{\Delta_{21}} & \frac{\lambda_{21} \Phi_3}{15\sqrt{2}}   \\
\frac{-{\lambda}_{21}\overline{v}_{2R}}{5\sqrt{6}} & \frac{-\sqrt{2}{\beta}_{12}\Phi_{3}}{\sqrt{15}} & 0 & \frac{\lambda_{21} \Phi_3}{15\sqrt{2}} & M_{\Delta}
\end{array}
\right),
\end{equation}

\begin{equation}\label{MT21}
B_{21}=\left(
\begin{array}{cccc}
 \frac{{\beta}_{21}{v}_{2R}}{\sqrt{5}} & 0&  -\frac{{\lambda}_{12}{v}_{2R}}{10\sqrt{3}}& -\frac{{\lambda}_{12}{v}_{2R}}{5\sqrt{6}}\\
 m_{H} & \beta_{12}\pht{H\Delta} & \overline{\beta}_{12}\pht{H\overline{\Delta}} & -\overline{\beta}_{12}\frac{\sqrt{2}\Phi_{3}}{\sqrt{15}} \\
 \overline{\beta}_{21}\pht{H{\Delta}} & {m}_{\Delta_{21}} & 0 & 0\\
  \beta_{21}\pht{H\overline{\Delta}} & 0 & {m}_{\Delta_{12}} & \frac{\lambda_{12} \Phi_3}{15\sqrt{2}}   \\
 -{\beta}_{21}\frac{\sqrt{2}\Phi_{3}}{\sqrt{15}} & 0 & \frac{\lambda_{12} \Phi_3}{15\sqrt{2}} & {m_{\Delta}}_{12}+\lambda_{12} \pht{\Delta}
\end{array}
\right),
\end{equation}
and
\begin{equation}\label{MT22}
B_{22}=\left(
\begin{array}{ccccc}
M_{\Phi} & \frac{{\beta}_{12}{v}_{1R}}{\sqrt{5}} & 0&  -\frac{{\lambda}_{21}{v}_{1R}}{10\sqrt{3}}& -\frac{{\lambda}_{21}{v}_{1R}}{5\sqrt{6}} \\
\frac{\overline{\beta}_{12}\overline{v}_{1R}}{\sqrt{5}} & \alpha_1 S & 0 & 0 &  0   \\
 0 & 0 & \alpha_2 S &  0  & 0 \\
 -\frac{{\lambda}_{12}\overline{v}_{1R}}{10\sqrt{3}} & 0 & 0 & \alpha_2 S &  0   \\
-\frac{{\lambda}_{12}\overline{v}_{1R}}{5\sqrt{6}} & 0 & 0 & 0 &  \alpha_2 S  \\
\end{array}
\right),
\end{equation}
where for simplicity we have defined
\begin{eqnarray}\label{doubelements}
\pht{H\Delta}&=&-\frac{\Phi_{1}}{\sqrt{10}}+\frac{\Phi_{2}}{\sqrt{30}}, \quad M_{\Delta}={m_{\Delta}}_{21}+\lambda_{21} \pht{\Delta},\nonumber\\
\pht{H\overline{\Delta}}&=&-\frac{\Phi_{1}}{\sqrt{10}}-\frac{\Phi_{2}}{\sqrt{30}},\quad M_{\Phi}=m_{\Phi}+\lambda(\frac{\Phi_1}{\sqrt{6}}+\frac{\Phi_2}{3\sqrt{2}}+\frac{2\Phi_3}{3}),\nonumber\\
\pht{\Delta}&=&\frac{\Phi_{1}}{10\sqrt{6}}+\frac{\Phi_{2}}{30\sqrt{2}}.\nonumber\\
\end{eqnarray}
The determinant of $M_T$ is nonzero and consequently $M_T$ is reversible with all
eigenvalues at GUT scale.

In SUSY GUTs, the dominant mechanism inducing proton decays is through the dimension-5 operators \cite{ETM1,ETM2}
\begin{equation}\label{dimension5}
-W_5=C_L^{ijkl}\frac{1}{2}q_iq_jq_kl_l+C_R^{ijkl}u_i^cd_j^cu_l^c e_k^c,
\end{equation}
which are called the $LLLL$ and $RRRR$ operators, respectively,
obtained by integrating out the  colored triplet Higgs superfields in the interactions in (\ref{superp1}).
The coefficients $C_L$s at the GUT scale $M_G$ are \cite{fukuyamageneral}
\begin{eqnarray}\label{cl}
C_L^{ijkl}(M_G)&=&Y^{ij}_{10}(M_T^{-1})_{11}Y^{kl}_{10}+Y^{ij}_{10}(M_T^{-1})_{12}Y^{kl}_{126}\nonumber\\
&+&Y^{ij}_{126}(M_T^{-1})_{31}Y^{kl}_{10}+Y^{ij}_{126}(M_T^{-1})_{32}Y^{kl}_{126}.
\end{eqnarray}
The Yukawa couplings have been rather constrained by fitting the fermion masses and mixing,
thus suppressing proton decay rates needs some detailed investigations on the matrix elements in $M_T$.

From (\ref{superp1}),
only  $H_1$ and $\overline{\Delta}_1$ couple with fermions, and hence it is the up-left $4\times4$ block of $M_T^{-1}$  that can affect the proton decay
through the dimension-5 operators.
These relevant elements in the $M_T^{-1}$ are proportional to
their corresponding algebraic complements divided by the determinant of $M_T$.
These corresponding algebraic complements are proportional to $\overline{v}_{1R}\sim M_I$ or $S_0\sim M_I$ in (\ref{MT22})
which is small compared with the GUT scale.
As a consequence,
the  proton decay amplitudes are suppressed by a factor $M_I/M_{G}$.

Now we have established in the present model a proportional relation  between the intermediate VEV $M_I$,
required by the seesaw mechanism, and the proton decay amplitudes.
Consequently,
the proton decay amplitudes are proportional to $\frac{M_I}{M_{G}^2}$,
substantially  suppressed compared to $\frac{1}{M_{G}}$ in the usual  models.
For the $RRRR$ type operators the results are just the same.

Relating the proton decay suppression with the seesaw VEV can be understood in other viewpoints.
Since only  part of the elements in the the up-left $4\times 4$ block
couple to the matter fields,
we can get a smaller effective mass matrix by integrating out the down-right $5\times 5$ block formally
\begin{equation}\label{effectivemass}
 M_{\textrm{eff}}=-B_{12}\cdot B_{22}^{-1} \cdot B_{21}.
\end{equation}
From (\ref{MT22}), $B_{22}$ has only one GUT scale mass eigenvalue.
Rotating the bases and transforming $B_{22}$ into diagonal form,
\begin{equation}\label{D22}
D_{22}=\textrm{diag}\ O\left(M_G, M_I, M_I, M_I, M_I \right),
\end{equation}
The elements of $B_{11}$ remain to be zero, while
those of $B_{12}$ and  $B_{21}$ are still of the order $O(M_G)$, i.e.,
\begin{equation}\label{MTp}
M_T\to \tilde{M}_T\simeq\left(
\begin{array}{cc}
0_{(4\times 4)} & M_{G(4\times 5)} \\
M_{G(5\times 4)}& D_{22(5\times 5)}
\end{array}
\right).
\end{equation}
Indeed, each one of the four $M_I$ eigenvalues in $B_{22}$ gives rise to an eigenvalue of the order $O(\frac{M_{G}^2}{M_I})$
 in $M_{\textrm{eff}}$.
The largest one, $M_G$, contributes as corrections of the order $O(M_G)$ to the above four eigenvalues
 in $M_{\textrm{eff}}$ and hence are negligible.
In summary, it is the lightest eigenvalue in $M_{\textrm{eff}}$ that dominates in proton decay, and it turns out to be
\begin{equation}\label{mhc}
M_{H_C}^{\textrm{eff}}\sim\frac{M_{G}^2}{M_I}\sim 2\times 10^{18}\textrm{GeV}.
\end{equation}
For general values of parameters of \0 GUTs,
it is definitely sufficient to suppress the proton decay rates to satisfy the current experimental limits.

This mechanism of suppression of proton decay can be equivalently achieved by another method.
The \textbf{210}-plet does not couple to the matter fields thus its color
triplet-antitriplet components can be integrated out first.
In result, the reduced mass matrix for the color triplet-antitriplet Higgs is now
$8\times 8$ whose four blocks are all $4\times 4$:
(i) $B_{11}$ keeps unchanged as a null matrix;
(ii) $B_{12}$ has its leftmost  column eliminated, while the other elements remain at $M_{G}$;
(iii) $B_{21}$ has its uppermost  row eliminated, while the other elements remain at  $M_{G}$;
(iv) $B_{22}$ has its leftmost  column and lowest  row eliminated,
while the other elements are the order $O(\frac{v_{1R}}{M_\Phi}\overline{v}_{1R})\sim M_I$.
In the limit $M_I\to 0$, this structure is an analogue to the mass matrix for the Higgs color triplets
in the flipped SU(5) model \cite{flipsu51,flipsu52,flipsu53,flipsu54} which,
as is well known,
has negligible contributions of dimension-5 operators to proton decay.
With the $M_I$ elements kept, the inducing suppressed proton decay amplitudes would be of the order $O(\frac{M_I}{M_G^2})$, same as (\ref{mhc}).


\section{The doublets}\label{doubletsec}
To get the almost massless MSSM doublets $H_u$ and $H_d$,
we need a minimal fine-tuning in the  mass matrix of the doublets.
In the present model, we have Higgs doublets as follows:
\begin{eqnarray}\label{phiud}
\varphi_{u}&=&(H_{1u},\Delta_{1u}, \overline{\Delta}_{1u}, \Phi_{u}, H_{2u},\Delta_{2u}, \overline{\Delta}_{2u}),\\
\varphi_{d}&=&(H_{1d}, \overline{\Delta}_{1d}, \Delta_{1d}, \Phi_{d},  H_{2d}, \overline{\Delta}_{2d}, \Delta_{2d}).
\end{eqnarray}
After symmetry breaking at the GUT scale, only one pair of Higgs doublets remain massless, i.e.,
\begin{equation}\label{phiud}
 H_{u} =\sum_{i=1}^7 \alpha_u^{i\ast} \varphi_{u}^i,\quad  H_{d} =\sum_{i=1}^7 \alpha_d^{i\ast} \varphi_{d}^i.
\end{equation}
The mass matrix for the doublets is symbolically written as
\begin{equation}\label{doublet}
M_D=\left(
\begin{array}{c|c}
0_{(4\times 3)} & M_{G(4\times 4)} \\
\hline
 M_{G(3\times 3)} &
 \begin{array}{cc} M_{I(3\times 4)} \end{array}
\end{array}
\right),
\end{equation}
whose determinant factorizes into the determinant of $M_{G(3\times 3)}$ times that of $M_{G(4\times 4)}$.
The existence of zero eigenvalue in $M_D$  thus requires the determinant of  either $M_{G(3\times 3)}$
or  $M_{G(4\times 4)}$ is zero.
In solving the eigenstates of $M_D$ the order $M_I$ entries can be taken as small perturbations.
The solutions corresponding to $\textrm{Det}(M_{G(4\times 4)})=0$ lead to small up-type quark masses which
is excluded by the heavy top quark mass.
The other solutions corresponding to $\textrm{Det}(M_{G(3\times 3)})=0$
give, up to normalization factors,
\begin{eqnarray}
\alpha_u^\ast&=&O(1,1,1,0,0,0,0),\label{phiu}\\
\alpha_d^\ast&=&O(\frac{M_I}{M_G},\frac{M_I}{M_G},\frac{M_I}{M_G},\frac{M_I}{M_G},1,1,1).
\end{eqnarray}
This just explains the large ratio of $\frac{m_t}{m_b}$,
and  further gives
\begin{equation}\label{mtmb}
\textrm{tan}\beta =\frac{v_u}{v_d} \approx \frac{m_t}{m_b}\frac{M_I}{M_G}\sim O(1),
\end{equation}
suggesting that a small $\textrm{tan}\beta$ is favored in the present model.
This indeed agrees with a similar result
\begin{eqnarray}\label{fits}
\frac{\alpha_u^1}{\alpha_d^1}\textrm{tan}\beta \sim 10^2, ~ \frac{\alpha_u^3}{\alpha_d^2}\textrm{tan}\beta\sim 10^2,
\end{eqnarray}
got by simply fitting the fermion parameters in SO(10) models from many groups \cite{msso10a2,fits2,fits3,fits4}
without considering other constraints.
In the present model, the ratios on the R.H.S. of (\ref{fits}), however, are predicted to be related to
the ratio $\frac{M_G}{M_I}$.
Also,  (\ref{phiu}) holds exactly, showing that there is no $\Phi_u$ component in $H_u$ which,
following \cite{ltj},
suggests that
it is the type-I instead of type-II seesaw mechanism that works in the present model.


\section{Comments and conclusion}\label{conclusion}
In this work we have proposed a SUSY SO(10) model for sufficient suppression of
proton decay.
The suppression is found to be linked with the intermediate VEV required
by the seesaw mechanism.
The seesaw mechanism turns out to be type-I.
Assuming that the two doublets in MSSM are achieved by fine-tuning,
we find the components of these doublets agree in magnitudes with those got by just fitting
the fermion masses and mixing.
Again, the ratios of components are linked to the ratio of the GUT scale versus the
intermediate VEV.
Since all the Higgs particles beyond the MSSM doublets are at GUT scale,
the unification of coupling constants will be  maintained by adjusting the
parameters.
Above the GUT scale, the gauge coupling of SO(10) will increase fast into the
non-perturbative region, as many of the SUSY SO(10) models do.
This, besides the required fine-tuning in the doublet sector,
is another unsatisfactory aspect of the model.

Extensions of the present model are straightforward.
More realistic SO(10) models usually require \textbf{120}-plet Higgs to fit the fermion sector \cite{msso10a2,fits2,fits3,fits4}.
By adding a pair of Higgs of \textbf{120}-plets with U(1) charges as $+1$ and $-1$, respectively,
none of the above conclusions fails.
The new prediction is $\frac{\alpha_u}{\alpha_d}\textrm{tan}\beta \sim 10^2$ for the
new components from \textbf{120}-plet which, again, agrees with the result by simply fitting the data \cite{msso10a2,fits2,fits3,fits4}.

Alternatively, if we use Higgs multiplets in \textbf{45}+\textbf{54} instead of \textbf{210}  to break SO(10),
proton decay can also be suppressed at the same level. However, since in this case, the \textbf{10} Higgs cannot couple with \textbf{126} or $\overline{\textbf{126}}$, to produce the correct contents of the doublets $H_u$ and $H_d$,
a pair of Higgs in \textbf{120}-plets are needed to be included at the beginning, because the \textbf{120}-plet can both couple with the \textbf{10}-plet and the $\textbf{126}/\overline{\textbf{126}}$ through \textbf{45}+\textbf{54}.



\begin{thebibliography}{99}

\bibitem{gut1}
J. C. Pati and A. Salam, \textit{Lepton number as the fourth ``color''}, \textit{Phys. Rev.} \textbf{D 10} (1974) 275.
\bibitem{gut2}
H. Georgi and S. L. Glashow, \textit{Unity of all elementary-particle forces}, \textit{Phys. Rev. Lett.} \textbf{32} (1974) 438.

\bibitem{seesawi1}
P.~Minkowski, $\mu\to e\gamma$ \textit{at a rate of one out of 109 muon decays}?, \textit{Phys.~Lett.}~\textbf{B~67}~(1977)~421.
\bibitem{seesawi2}
T.~Yanagida, in \textit{workshop on unified theories}, KEK Report 79-18~(1979)~95.
\bibitem{seesawi3}
M. Gell-Mann, P. Ramond, and R. Slansky, in  \textit{Supergravity}, P. van Nieuwenhuizen and D.Z. Freedman (eds.), North Holland Publ. Co., (1979) p. 315.
\bibitem{seesawi4}
S. L. Glashow, in \textit{1979 Cargese Summer Institute on Quarks and Leptons}, Plenum, New York, (1980) p. 687.
\bibitem{seesawi5}
R. N. Mohapatra and G. Senjanovi$\acute{\textrm{c}}$, \textit{Neutrino mass and spontaneous parity nonconservation}, \textit{Phys. Rev. Lett.} \textbf{44} (1980) 912.

\bibitem{seesawii1}
G. Lazarides, Q. Shafi, and C. Wetterich, \textit{Proton lifetime and fermion masses in an} SO(10) \textit{model}, \textit{Nucl. Phys.} \textbf{B 181} (1981) 287.
\bibitem{seesawii2}
R. N. Mohapatra and G. Senjanovi$\acute{\textrm{c}}$, \textit{Neutrino masses and mixings in gauge models with spontaneous parity violation}, \textit{Phys. Rev.} \textbf{D 23} (1981) 165.
\bibitem{seesawii3}
J. Schechter and J.W. F. Valle, \textit{Neutrino masses in} SU(2)$\otimes$U(1) \textit{theories}, \textit{Phys. Rev.} \textbf{D 22} (1980) 2227.
\bibitem{seesawii4}
E. Ma and U. Sarkar, \textit{Neutrino masses and leptogenesis with heavy Higgs triplets}, \textit{Phys. Rev. Lett.} \textbf{80} (1998) 5716.

\bibitem{so101}
H. Georgi, in \textit{Proceedings of APS Division of Particles and Fields}, edited by C. Carlson, (1975) p. 575.
\bibitem{so102}
H. Frtzsch and P. Mikowski,  \textit{Unified interactions of leptons and hadrons}, \textit{Ann. Phys.} \textbf{93} (1975) 193.

\bibitem{msso10a1}
C. S. Aulakh, B. Bajc, A. Melfo, G. Senjanovi$\acute{\textrm{c}}$, and F. Vissani, \textit{The minimal supersymmetric grand unified theory}, \textit{Phys. Lett.} \textbf{B 588} (2004) 196.
\bibitem{msso10a2}
T. Fukuyama, T. Kikuchi, A. Ilakovac, S. Meljanac, and N. Okada, \textit{Detailed analysis of proton decay rate in the minimal supersymmetric} SO(10) \textit{model}, \textit{JHEP} 0409 (2004) 052.

\bibitem{msso10b}
B. Bajc, A. Melfo, G. Senjanovi$\acute{\textrm{c}}$, and F. Vissani, \textit{Minimal supersymmetric grand unified theory: Symmetry breaking and the particle spectrum}, \textit{Phys. Rev.} \textbf{D 70} (2004) 035007.

\bibitem{nmsso101}
T. Fukuyama, A. Ilakovac,  T. Kikuchi, S. Meljanac, and N. Okada,  SO(10) \textit{Group theory for the unified model building},  \textit{J. Math. Phys.} \textbf{46} (2005)  033505.
\bibitem{nmsso102}
M. Malinsk$\acute{\textrm{y}}$, arXiv:0807.0591.

\bibitem{dlz}
L. Du, X. Li and D.-X. Zhang, \textit{Proton decay in a supersymmetric} SO(10) \textit{model with missing partner mechanism}, \textit{JHEP} 04(2014) 027.

\bibitem{moha}
B. Dutta, Y. Mimura, and R.N. Mohapatra, \textit{Suppressing proton decay in the minimal} SO(10) \textit{model}, \textit{Phys. Rev. Lett.} \textbf{94} (2005) 091804.

\bibitem{dts1}
S. Dimopoulos and F. Wilczek, NSF-ITP-82-07.
\bibitem{dts2}
K. S. Babu, I. Gogoladze and Z. Tavartkiladze, \textit{Missing partner mechanism in} SO(10) \textit{grand unification}, \textit{Phys. Lett.} \textbf{B 650} (2007) 49.

\bibitem{ETM1}
N. Sakai and T. Yanagida, \textit{Proton decay in a class of supersymmetric grand unified models}, \textit{Nucl. Phys.} \textbf{B 197} (1982) 533.
\bibitem{ETM2}
S. Weinberg, \textit{Supersymmetry at ordinary energies. Masses and conservation laws}, \textit{Phys. Rev.} \textbf{D 26} (1982) 287.

\bibitem{green1}
M. B. Green and J. H. Schwarz, \textit{Anomaly cancellations in supersymmetric $D=10$ gauge theory and superstring theory}, \textit{Phys. Lett.} \textbf{B 149} (1984) 117.
\bibitem{green2}
M. Dine, N. Seiberg, and E. Witten, \textit{Fayet-Iliopoulos terms in string theory}, \textit{Nucl. Phys.} \textbf{B 289} (1987) 589.
\bibitem{green3}
J. J. Atick, L. J. Dixon, and A. Sen, \textit{String calculation of fayet-iliopoulos D-terms in arbitrary supersymmetric compactifications}, \textit{Nucl. Phys.} \textbf{B 292} (1987) 109.
\bibitem{green4}
M. Dine, I. Ichinose, and N. Seiberg, \textit{F terms and D terms in string theory}, \textit{Nucl. Phys.} \textbf{B 293} (1987) 253.

\bibitem{fukuyamageneral}
T. Fukuyama, A. Ilakovac, T. Kikuchi, S. Meljanac, and N. Okada, \textit{General formulation for proton decay rate in minimal supersymmetric} SO(10) \textit{GUT}, \textit{Eur. Phys. J.} \textbf{C 42} (2005) 191.

\bibitem{flipsu51}
S. M. Barr, \textit{A new symmetry breaking pattern for} SO(10) \textit{and proton decay}, \textit{Phys. Lett.} \textbf{B 112} (1982) 219.
\bibitem{flipsu52}
A. De Rujula, H. Georgi, and  S.L. Glashow, \textit{Flavor goniometry by proton decay}, \textit{Phys. Rev. Lett.} \textbf{45} (1980) 413.
\bibitem{flipsu53}
J.P. Derendinger, J. E. Kim, and D. V. Nanopoulos, \textit{Anti}-Su(5), \textit{Phys. Lett.} \textbf{B 139} (1984) 170.
\bibitem{flipsu54}
I. Antoniadis, J. R. Ellis, J. S. Hagelin, and D. V. Nanopoulos, \textit{Supersymmetric flipped} SU(5) \textit{revitalized}, \textit{Phys. Lett.} \textbf{B 194} (1987) 231.

\bibitem{fits2}
W. Grimus and H. K$\ddot{\textrm{u}}$hb$\ddot{\textrm{o}}$ck, \textit{Fermion masses and mixings in a renormalizable} SO(10)$\times$Z$_2$ \textit{GUT}, \textit{Phys. Lett.} \textbf{B 643} (2006) 182.
\bibitem{fits3}
G. Altarelli and G. Blankenburg, \textit{Different} SO(10) \textit{paths to fermion masses and mixings}, \textit{JHEP} 1103 (2011) 133.
\bibitem{fits4}
A. Dueck and W. Rodejohann, \textit{Fits to} SO(10) \textit{grand unified models}, \textit{JHEP} 1309 (2013) 024.

\bibitem{ltj}
J. Kang, P. Langacker, and T.-j. Li, \textit{Neutrino masses in supersymmetric} SU(3)$_C\times$SU(2)$_L\times$U(1)$_Y\times$U(1)$^\prime$ \textit{models}, \textit{Phys. Rev.} \textbf{D 71} (2005) 015012.

\end{thebibliography}
\end{document}